\documentclass{aa}
\usepackage{graphicx}

\def\mdot{\mbox{\,$\dot{M}$}}
\def\msunyr{\mbox{\,$M_{\odot}\,{\rm yr^{-1}}$}}

\def\lessim{\mathrel{\hbox{\rlap{\hbox{\lower4pt\hbox{$\sim$}}}\hbox{$<$}}}}
\def\gtrsim{\mathrel{\hbox{\rlap{\hbox{\lower4pt\hbox{$\sim$}}}\hbox{$>$}}}}

\begin{document}

   \title{The Late-Time Radio spectrum of SN1993J}


   \author{M.A.\ P\'erez-Torres 
          \inst{1}
          \and A.\ Alberdi\inst{2}
          \and J.M.\ Marcaide\inst{3}
   }

   \offprints{M.A.\ P\'erez-Torres, \email{torres@ira.cnr.it}}

   \institute{Istituto di Radioastronomia,
              Via P. Gobetti 101, 40129 Bologna, Italy \\
           email: torres@ira.cnr.it 
         \and
	   Instituto de Astrof\'{\i}sica de Andaluc\'{\i}a, CSIC, Apdo.
	   Correos 3004, E-18080 Granada, Spain\\
           email: antxon@iaa.es
        \and 
	   Departamento de Astronom\'{\i}a y 
           Astrof\'{\i}sica, Universidad de Valencia, 
	   46100 Burjassot, Spain
	   email: J.M.Marcaide@uv.es
   }

\date{A\&A Main Journal, in press}

\abstract{
We present VLA radio continuum measurements of SN1993J in M81
at the frequencies of 0.32 (P-band), 1.3 and 1.7 (L-band),
4.9 (C-band), 8.5 (X-band), and 14.9 (U-band) GHz
carried out on December 17 and 21, 2000, about 2820 days after 
the supernova explosion.
We find that a power-law spectrum, free-free absorbed by an homogeneous,  
or clumpy, distribution of ionized gas yields the best fit to 
the radio data. 
A combined homogeneous-clumpy model is not favored, but neither 
totally excluded.
This result contrasts with the modeling 
of the early ($t\,\lessim\,$230 days) radio emission from SN1993J,
where a mixture of homogeneous and clumpy absorbers appeared to be
necessary to adequately describe the behavior of the light curves. 
The radio spectrum of supernova SN1993J between 0.32 and 14.9 GHz
is well characterized by $\alpha = -0.67 \pm 0.02 $ ($S_\nu \propto \nu^\alpha$),
typical of an optically thin radio supernova.
A fit to the radio spectra of SN1993J from $\sim$ 70 up to 2820 days shows
that the observed spectral index of SN1993J
has been slowly evolving since $t\sim$1000 days on, with 
the observed spectral index changing from 
$\alpha \approx -1$  to $\alpha=-0.67$. 
This spectral evolution seems to suggest that, in addition to the 
radiative (synchrotron) losses,  
adiabatic cooling and ionization (Coulomb) losses at the lowest frequencies 
might be contributing significantly to the integrated electron spectrum. 
}

\maketitle

\titlerunning{The late radio spectrum of SN1993J}
\authorrunning{P\'erez-Torres et al.}

\section{Introduction}
\label{sec:intro}
The most widely accepted model for the radio emission from 
supernovae is the standard interaction model, in which the observed 
radio (synchrotron) emission is associated with the interaction region between the 
supernova ejecta and the pre-supernova wind of the progenitor star. 
The radio emission shows an early rise and 
a late decline due to a low-frequency absorption process. 
Chevalier (\cite{che82}) considered a number of mechanisms that could 
account for the absorption: 
(i) free-free absorption by material in front of the forward shock front, 
(ii) free-free absorption by material within the emitting region, 
(iii) synchrotron self-absorption, and 
(iv) the Razin-Tsytovich effect. 
At the beginning of the 1980's, the available radio light curves could
be well explained by synchrotron radio emission that was 
partially free-free absorbed by ionized thermal material in 
the circumstellar wind of the progenitor star (e.g. Weiler et al.~\cite{wei86}). 
The availability of more complete light curves for these supernovae
(e.g. Weiler et al.~\cite{wei89}, \cite{wei90}) 
provided further support for this mechanism. 
However, as new radio 
supernovae have been detected, the situation has become more complex.
For example, the radio light curves of SN1986J showed a 
different shape in the rise to maximum, which was interpreted by Weiler et al. (\cite{wei90}) 
as evidence for internal free-free absorption.  
These authors suggested that a filamentary, clumpy-like structure of thermal absorbers 
could also be the source of the slow radio turn-on. 
Chevalier (\cite{che98}) has recently argued that synchrotron self-absorption 
is likely to be the dominant absorption mechanism for some radio supernovae, 
in particular for SN1987A, and for Type Ib and Type Ic supernovae.
Finally, observations of SN1993J seem to indicate that more than one 
absorption mechanism plays a role 
(Fransson and Bj\"ornsson \cite{fb98}; P\'erez-Torres et al.~\cite{pam01}; 
Mioduszewski et al.~\cite{mio01}).

 SN1993J is to date the radio supernova whose evolution has been monitored 
in greatest detail. 
The brightness and proximity of SN1993J in M81
offered an unprecedented occasion for VLBI studies, 
almost since the supernova explosion on 28 March 1993.
For the first time, a shell-like structure in a young
radio supernova was discovered (Marcaide et al. \cite{mar95a}), in
agreement with the circumstellar interaction model.
Marcaide et al. (\cite{mar95b}) showed that the expansion of SN1993J was
self-similar, and produced the first movie of an expanding supernova.
Later on, Marcaide et al. (\cite{mar97}) reported on the deceleration
of the SN1993J expansion, a result recently confirmed by Bartel et al. 
(\cite{bartel00}). 
The modeling of the radio light curves of SN1993J 
has also given deeper insight into the supernova phenomenon. 
It seems now clear that the progenitor of SN1993J had a substantial 
mass-loss rate during the late stages of its evolution (\mdot $\ge 10^{-5}$ 
\msunyr,  e.g. Van Dyk et al. \cite{vdyk94}). 
The power-law density profile 
of the circumstellar medium around SN1993J seems to be different from the standard one 
(Lundqvist \cite{lun94}; Van Dyk et al. \cite{vdyk94};
Marcaide et al. \cite{mar97}; Immler et al. \cite{immler01};
Mioduszewski et al.~\cite{mio01}), although 
Fransson and Bj\"ornsson (\cite{fb98}) have succeeded in modeling 
the SN1993J radio light curves with an $s$=2 density profile for the CSM. 

Based on numerical modelings of the available radio 
data for SN1993J, Fransson and Bj\"ornsson (\cite{fb98}) and 
P\'erez-Torres et al. (\cite{pam01}) made predictions about its 
late-time, low frequency ($< 1.4$ GHz) radio light behaviour. 
However, observational results of the low-frequency radio emission from SN1993J were 
lacking.
Thus, we carried out VLA observations between $\sim$322  and $\sim$15000 MHz 
aimed at checking the above modeling efforts,
as well as a way to
discern the main absorbing processes acting in SN1993J. 

\section{Observations and Data Reduction}
\label{sec:obs}

We observed SN1993J using 25 antennas of the VLA in 
its most extended configuration 
(A, with baseline lengths from 0.68 up to 36.4 km)
at L (18\,cm) and P (90\,cm) bands 
from UT 13:55 to UT 15:25 on 17 December 2000, 
and at U (14.9 GHz), X(8.5 GHz), C(4.9 GHz), L(1.3-1.7 GHz), and P(0.32 GHz) bands 
from UT 08:40 to UT 11:40 on 21 December 2000.
Each frequency band was split into two intermediate frequencies (IFs).  
Data processing was made using the Astronomical Image
Processing System ({\it AIPS}).  
We first describe briefly the observations at the higher frequencies, 
and then emphasize and describe in detail those at lower 
frequencies, particularly those in the P-band, 
deemed to be a discriminator of the dominant 
absorption mechanism in SN1993J.
Table \ref{tab:vla_log} summarizes our observations. 
The flux densities reported were obtained by combining the data from 
both intermediate frequencies (IFs) at each frequency band, 
except at L-band, for which the frequency separation was large enough 
that we could use each flux density value independently. 

\subsection{High-frequency (U-, X-, and C-band) observations}
We used a standard continuum mode at U, X, and C bands, thus covering a 
bandwidth of 50 MHz per IF and frequency band. At all three bands,
we observed SN1993J phase-referenced to 0917+624. The source 
0917+624 also served as the phase reference for the system calibration.
At U- and X-band, 3C286 was used as primary flux calibrator (assumed of constant 
flux density), while at C-band 3C147 was used instead.
At U-band, the integration times on SN1993J
and  0917+624 were $\tau_{\rm 93J}$=21 min and $\tau_{\rm 0917}$ = 7 min, 
while at X-band  these were 11.5 and 6 min, and at C-band 12 and 7 min, respectively.

\subsection{Low-frequency (L- and P-band) observations}

Observations were done in spectral-line mode in order to detect the 
likely presence of radio frequency interference (RFI) in the data, and 
also to avoid bandwidth smearing. 
In addition, this observing mode allows a more careful editing. 
3C286 was used at both bands as primary flux calibrator, 
as well as bandpass calibrator.

{\it L-band observations}.
Each of the two 12.5-MHz IFs at L-band (centered at 1.34 and 1.67 MHz, 
respectively) were split into 8 spectral channels.
Observations of SN1993J ($\tau_{\rm 93J}$ = 11.5 min) 
were phase-referenced to 0917+624 ($\tau_{\rm 0917}$ = 10.5 min).
0917+624 was also used as secondary flux calibrator at L-band.  
Our final hybrid map at L-band (not shown here)   
was made at an effective observing frequency of 1.50 GHz.
The peak of brightness of the map (97.4 mJy beam$^{-1}$) 
corresponds to the nucleus of M81, and the flux density 
of SN1993J is 31.9 mJy. 
The rms of the image background is 0.3 mJy/beam.  
The flux density values reported 
at L-band in Table \ref{tab:vla_log} correspond 
to each IF, 1.34 and 1.67 GHz, respectively. 
Since the center frequencies in L-band are quite separated, 
they can be used as independent values for the purpose of fitting a 
spectral index (see below).  We note that, though the distance between SN1993J 
and M81 (2$\arcmin\,48\arcsec$) is much larger than the dimensions of the restoring beam 
($1.84\times0.99$ arcsec), wide field mapping is necessary
if one aims at accurate determinations of the flux densities of SN1993J 
at low frequencies.

{\it P-band observations}.
At P-band, each of the two 3.125-MHz IFs (centered at 322 and 327 MHz, 
respectively) was split into 32 spectral channels, which permitted a 
detailed inspection of the data, looking for the presence of RFI.
Since 0917+624 is a relatively weak source at P-band, unlike at 
higher frequencies, confusion of many strong sources in the field made it 
unsuitable as secondary calibrator. Therefore, the much stronger source 
J1206+6414 (3C268.3) was used instead. The total integration times 
were $\tau_{\rm 93J}\,\approx$78 min, and $\tau_{\rm 3C268.3}\,\approx$39 min.
We imaged 3C268.3 to obtain an accurate value of its flux density.
We noticed that the coordinates used for 3C268.3 at the correlator
($\alpha = 12^{\rm h}\, 06^{\rm m}\, 24\fs711$, 
$\delta = +64\degr\, 14\arcmin\, 40\farcs939$; 
J2000.0) were off from the actual ones by more than one arcmin.  Indeed, 
the coordinates obtained using the {\it AIPS} task JMFIT were $\alpha = 
12^{\rm h}\, 06^{\rm m}\, 24\fs8978$, 
$\delta = +64\degr\, 13\arcmin\, 36\farcs425$ (J2000.0). 
Such position offset for 3C268.3 translated into a similar one for
SN~1993J and the whole sky around it. 
We corrected the phases of 3C268.3 for this offset.

\begin{figure*}[htb!]
 \label{fig:wide_field}
 \includegraphics[angle=0,width=\textwidth]{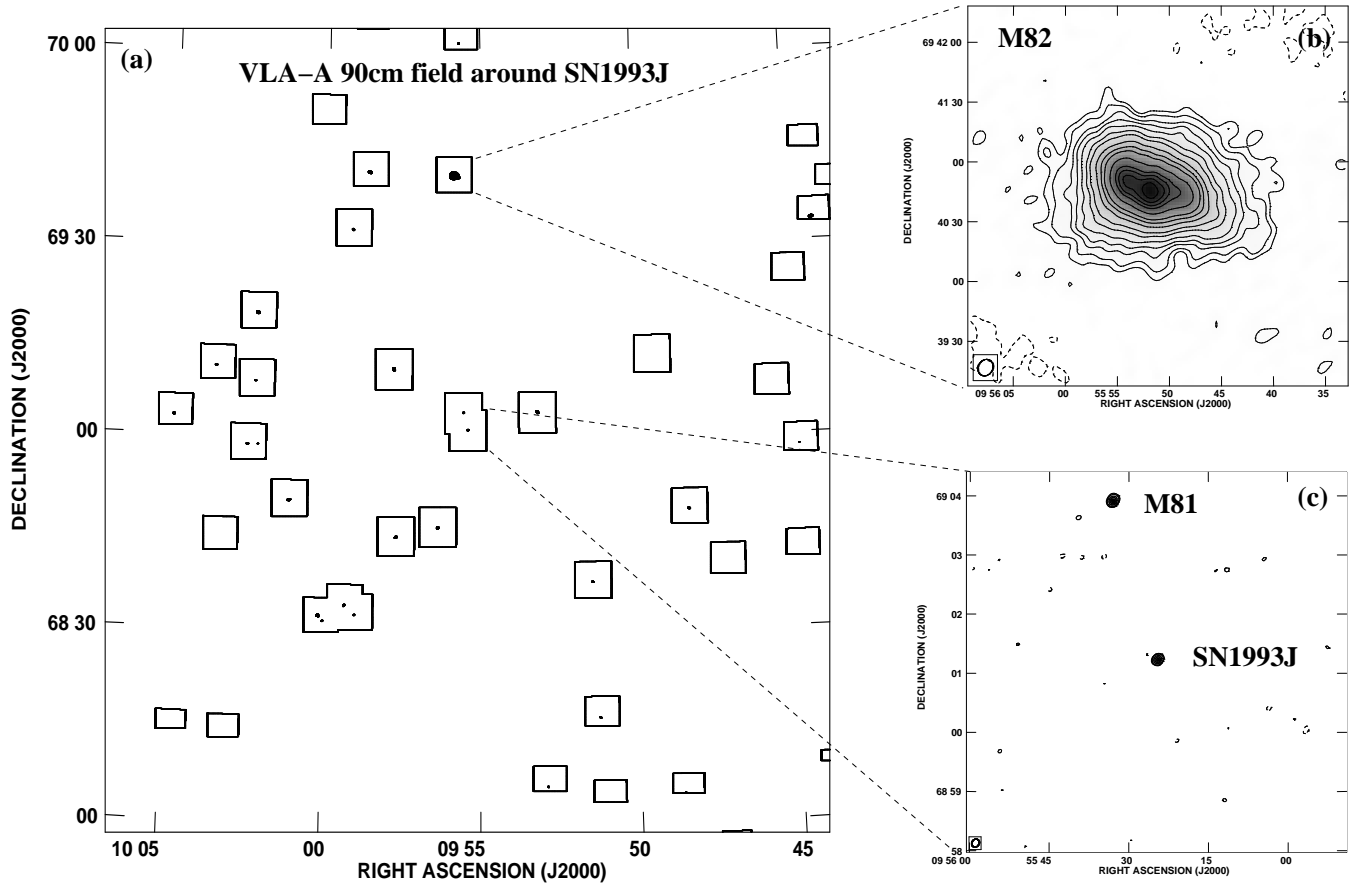}
 \caption{ 
{\bf (a)} Field of of view of the VLA in A configuration (VLA-A) around SN1993J 
at P-band, on 17 and 21 December, 2000.
The image shows only the inner 2$\times$2 deg (rather 
than the 4$\times$4 deg actually mapped),  which includes
38 fields out of the 78 used in the hybrid mapping process.
The image is centered on SN1993J. 
Only contours above 20 mJy beam$^{-1}$ have been drawn. 
{\bf (b)} Hybrid map of the galaxy M~82.
The contours are (-2,2,3,5,7,10,15,20,30,40,50,70,90,100,110,120) $\times$ 
4.2 mJy beam$^{-1}$, the off-source rms noise. 
The peak of brightness of the map 
($\alpha = 09^{\rm h}\, 55^{\rm m}\,51\fs814$, 
$\delta = +69\degr\,40\arcmin\, 45\farcs20$; J2000.0) 
is 537.8 mJy beam$^{-1}$. The flux density in the image above the 2-rms level is 10.49 Jy.
The dimensions of the restoring beam
are $8.6\times7.2$ arcsec, with the beam's major axis
oriented along P.A.=-30.2 deg.
{\bf (c)} Hybrid map of SN1993J and the nucleus of its host galaxy M81. 
The contours are (-3,3,5,10,15,20,25,30) $\times$ 3.3 mJy beam$^{-1}$,
the off-source rms noise. 
The peak of brightness of the map corresponds to the supernova
and is 71.1 mJy\,beam$^{-1}$. 
The dimensions of the restoring beam
are $8.6\times7.2$ arcsec, with the beam's major axis
oriented along P.A.=-30.9 deg.
}
\end{figure*}

With the 3C268.3 data corrected, we imaged 
SN1993J at P-band, aimed to determine its flux density. 
Since the primary beam of the VLA at P-band is so large 
($\approx2\degr\,20\arcmin$), bright sources far from 
the supernova could greatly affect our image.
Further, to obtain a good signal-to-noise ratio (SNR) in the image, one 
is forced to map even distant, strong sources. 
The situation is similar to that encountered 
at L-band, where self-calibration was mandatory to improve the SNR 
of the image, but useless unless M81 was also included in the imaging process. 
At P-band, not only the M81 nucleus is important, 
but also far away, strong sources such as 
M82 need also to be taken into account to be able to apply self-calibration
to the data.  
We used the {\it AIPS} task SETFC to create a list of fields with 
suspected sources in them, rather than making an enormous, 
computationally unmanageable image. It is worth noticing
that the interferometric phases are initially so poorly calibrated 
that the reconstructed image will show very few sources, making 
very difficult the convergence of the hybrid mapping procedure. 
Therefore, we used the {\it AIPS} procedure FACES,
which searchs for sources in the NVSS catalog, 
to supply an initial model of the sky, 
and thus improve the global phase calibration.  
We then applied wide-field imaging on this self-calibrated dataset:
since the array configuration looks different to sources in different
parts of the (large) primary beam, we handled this by 
computing the uv-coverage differently for many small fields 
within the task IMAGR. 
Sources above a threshold of 33 mJy (at a frequency of 1.4 GHz) 
were searched for as ``a priori''  candidates in the NVSS catalog, 
within a 2$\times$2 deg region around SN1993J.  
This resulted in 78 fields that were 
cleaned  using wide-field mapping. 
Fig. \ref{fig:wide_field} shows 
the inner 1$\times$1 deg of our final, composite image of
the VLA-A field at P-band.
For clarity, only contours above 20 mJy\,beam$^{-1}$ have been drawn.
Panel (b) shows the galaxy M~82, 
offset about $0\farcm40$ from SN1993J.  
The latter is shown in panel (c) along with the nucleus of its host galaxy M81. 
M82 has a flux of $\sim$10.5 Jy, at an epoch when SN1993J has a flux of a mere
71.1 mJy (see Table~\ref{tab:vla_log} for details of the observations).

\section{The radio spectrum of SN1993J}
\label{results}

\begin{table*}
 \caption[]{VLA Observing Results}
  \label{tab:vla_log}
$$
 \begin{array}{lrrrrrr}
   \hline\noalign{\smallskip}
  \multicolumn{1}{l}{\rm Source\, Name} & 
  \multicolumn{5}{c}{\rm Flux\, density\,(mJy) } \\ 
	\cline{2-7}\noalign{\smallskip} 
&
\multicolumn{1}{c}{14.94\, \rm GHz} &
\multicolumn{1}{c}{8.46\, \rm GHz} &
\multicolumn{1}{c}{4.86\, \rm GHz} &
\multicolumn{1}{c}{1.67\, \rm GHz} &
\multicolumn{1}{c}{1.34\, \rm GHz} &
\multicolumn{1}{c}{0.324\, \rm GHz}  \\
   \hline\noalign{\smallskip}
\multicolumn{1}{l}{\rm SN1993J} &
\multicolumn{1}{c}{6.7 \pm 0.1} &
\multicolumn{1}{c}{10.4 \pm 0.2 } &
\multicolumn{1}{c}{14.7 \pm 0.4 } &
\multicolumn{1}{c}{30.5 \pm 0.4} &
\multicolumn{1}{c}{33.7 \pm 0.4} &
\multicolumn{1}{c}{71.1  \pm 3.4} \\ 
\multicolumn{1}{l}{\rm 3C286^{(1)}} &
\multicolumn{1}{c}{3293 \pm 16} &
\multicolumn{1}{c}{4939 \pm 5} &
\multicolumn{1}{c}{-} &
\multicolumn{1}{c}{13708 \pm 20} &
\multicolumn{1}{c}{15157 \pm 21} &
\multicolumn{1}{c}{25940 \pm 120} \\ 
\multicolumn{1}{l}{\rm 0917+624^{(2)}} &
\multicolumn{1}{c}{1164 \pm 1} &
\multicolumn{1}{c}{1370 \pm 1}  & 
\multicolumn{1}{c}{1472 \pm 1}  &
\multicolumn{1}{c}{1272 \pm 2}  &
\multicolumn{1}{c}{1194 \pm 3}  &
\multicolumn{1}{c}{-} \\ 
\multicolumn{1}{l}{\rm 3C147^{(1)}} &
\multicolumn{1}{c}{ - } &
\multicolumn{1}{c}{ - } &
\multicolumn{1}{c}{7940 \pm 10} &
\multicolumn{1}{c}{ - } &
\multicolumn{1}{c}{ - } &
\multicolumn{1}{c}{25940 \pm 120} \\ 
\multicolumn{1}{l}{\rm 3C268.3^{(2)}} &
\multicolumn{1}{c}{ - } &
\multicolumn{1}{c}{ - } &
\multicolumn{1}{c}{ - } &
\multicolumn{1}{c}{ - } &
\multicolumn{1}{c}{ - } &
\multicolumn{1}{c}{10040 \pm 20} \\ 
\hline\noalign{\smallskip}
 \end{array}
$$
\begin{list}{}{}
\item[] {\rm $^{(1)}$ primary flux calibrator, $^{(2)}$ phase and secondary flux calibrator. 
         All flux density values reported here were obtained by imaging each source, 
	 and the quoted uncertainty for each flux value corresponds to the off-source rms 
	 noise in each image.
         }
\end{list}

\end{table*}

\begin{figure}[htb!]
 \centering
 \includegraphics[angle=-90,width=9cm]{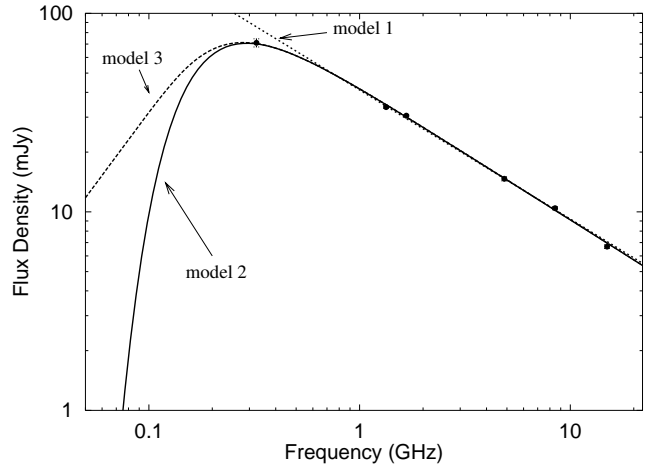}
 \caption{
Fits to the continuum radio spectrum of SN1993J on 17 and 21 December 2000. 
As indicated in the figure, the dotted, solid, and dashed lines correspond to 
models 1, 2, and 3, respectively, as described in the text. 
Note that the size of each data point but the one at 0.324 GHz, is larger than
its associated error bar.
}
 \label{fig:fits}
\end{figure}

We fitted the radio spectrum of SN1993J obtained
on 17 and 21 December 2000 using three models for its radio emission. 
These models are described by the equations

\begin{equation}
   S(\nu) = K_1\,\nu_5^{\alpha} 
\end{equation}

\begin{equation}
   S(\nu) = K_1\,\nu_5^{\alpha}\,e^{-\tau}
\end{equation}

\begin{equation}
 S(\nu) =K_1\, \nu_5^\alpha \, (1 - e^{-\tau^\prime})/\tau^\prime
\end{equation}

where $\tau   = K_2\,\nu_5^{-2.1}$, and $\tau^\prime = K_3\,\nu_5^{-2.1}$.
These equations represent a simple power-law spectrum (model 1), 
a power-law spectrum free-free absorbed by a screen of homogeneously 
distributed ionized gas (model 2), 
and 
a power-law spectrum free-free absorbed by a ``clumpy'' medium (model 3).  
To facilitate comparisons with the modeling of Van Dyk et al. (\cite{vdyk94}),  
we have used here their notation: 
$\nu_5$ is the frequency, in units of 5 GHz; 
$\alpha$ is the observed spectral index; 
$K_1$ is the flux density at 5 GHz, in mJy; 
and $K_2$ and $K_3$ are the optical depths at 5 GHz of the
homogeneous and ``clumpy'' external absorbing media, respectively, 
at the epoch of our observations.
Table \ref{tab:fits} summarizes our results, and Fig. \ref{fig:fits} 
shows the best fits to models 1 through 3. 
A simple power-law spectrum (model 1) yields a good fit to the data
from 14.94 GHz down to 1.34 GHz, but departs significantly from the
actual flux density at frequencies below 400 MHz, and can 
be ruled out.  
The best fits are given by models 2 (homogeneous absorbing 
medium) and 3 (clumpy absorbing medium), which yield an observed
spectral index $\alpha = -0.67\pm0.02$. 
The frequency at which the free-free optical depth becomes unity
is very similar for both model 2 ($\nu_{\tau =1} = 169$ MHz) and 
model 3 ($\nu_{\tau =1} = 241$ MHz). 
Note, however, that the very-low radio frequency behavior of the spectrum 
is significantly different (see Fig. \ref{fig:fits}). 
While model 2 (solid line) predicts a sharp fall-off below frequencies 
of $\sim$200 MHz, model 3 (dashed line) predicts 
a much milder flux density decrease. 
Unfortunately, the absence of radio measurements below 322 MHz 
does not allow us to rule out either of these models. 
\begin{table*}
 \caption[]{Radio spectrum fits for the VLA observations of SN1993J on 17 and 21 December 2000}
 \label{tab:fits}
$$
 \begin{array}{lccccccr}
  \hline
  \noalign{\smallskip}
  {\rm Model } & \alpha & K_1 {\rm (mJy)} & K_2\times10^4 & \nu_{\rm c,2} {\rm (MHz)} & 
                                K_3\times10^4 & \nu_{\rm c,3} {\rm (MHz)} & \chi_{\rm red}^2 \\
  \noalign{\smallskip}
  \hline
  \noalign{\smallskip}
      1 & 
           -0.65\pm0.02  & 14.4\pm0.4 & - & - & - & - & 8.9 \\
      2 & 
           -0.67\pm0.02 &  14.4\pm0.2 & 8\pm3 & 169^{+31}_{-38} & - & - & 3.6 \\
      3 & 
           -0.67\pm0.02 &  14.4\pm0.2 &  -  & - & 17\pm7 & 241^{+42}_{-52} & 3.6\\
  \noalign{\smallskip}
  \hline
 \end{array}
$$
\begin{list}{}{}
\item[]  $\alpha$ is the observed spectral index; $K_1$ is the flux density at 
 a frequency of 5 GHz; $K_2$ and $K_3$ are the optical depths at 5 GHz of 
the homogeneous and ``clumpy'' external absorbing media, respectively; 
$\chi^2_{\rm red}$ is the $\chi^2$ per degree of freedom.
Models 1 through 3 are explained in the text.
($\nu_{\rm c,2}$ and $\nu_{\rm c,3}$ are not fitted, and correspond
to the frequencies at which the free-free optical depth 
is unity for the homogeneous and ``clumpy'' medium, respectively.)
\end{list}{}{}
\end{table*}

Thus, from Table \ref{tab:fits} we cannot exclude, or prefer, any of the
two alternative models.
 
To model the early-time ($t\lessim230$ days) radio light curves 
of SN1993J, 
Van Dyk et al. (\cite{vdyk94}) invoked a combined 
``homogeneous-clumpy'' model of the form 
$ S_\nu \propto \nu_5^\alpha \, e^{-\tau}\, (1 - e^{-\tau^\prime})/\tau^\prime$. 
For such model, we can also obtain acceptable solutions 
(e.g. $\alpha=-0.67; K_1=14.4; K_2=8\times\,10^{-4}; K_3=10^{-5}$) 
at the expense of increasing $\chi^2_{\rm red}$ by $\approx\,50\%$. 
If the fitting parameters are left unconstrained, this model yields an even
better fit than models 2 and 3, but at the unaffordable price of a 
negative --hence unphysical-- value for $K_2$, or $K_3$.
Clearly, the combined ''homogeneous-clumpy'' model cannot be preferred over
models 2 or 3, but we also feel we cannot exclude it
solely on the basis of a 50\% increase of the reduced $\chi^2$.

For a power-law relativistic electron distribution, $N_E=N_0\,E^{-\gamma}$, 
the  spectral index of the electron distribution $\gamma$
is related to the observed spectral index: $\gamma = 1 - 2\,\alpha$. 
From our value of $\alpha$, it then follows that $\gamma=2.34\pm0.04$.
This value of $\gamma$ agrees well with that predicted by Fransson \& 
Bj\"ornsson (\cite{fb98}) for $t \gtrsim 1000$ days.
We should note that synchrotron self-absorption could play a role in the 
radio emission from SN1993J (Chevalier \cite{che98}, Fransson \& Bj\"ornsson 
\cite{fb98}, P\'erez-Torres et al. \cite{pam01}, Mioduszewski et al. \cite{mio01}).
In particular, the modeling of Fransson \& Bj\"ornsson (see their Fig. 12), 
predicted a flux density of $\sim 70$ mJy at 0.32 GHz and for 
the approximate epoch of our observations, while the modeling of P\'erez-Torres et al. predicted 
at that frequency a flux density larger than 100 mJy.
Such an excess in flux of the latter modeling is likely related to 
the different spectral evolution obtained for the electron spectrum.

\begin{figure}[htb!]
 \includegraphics[angle=-90,width=9.0cm]{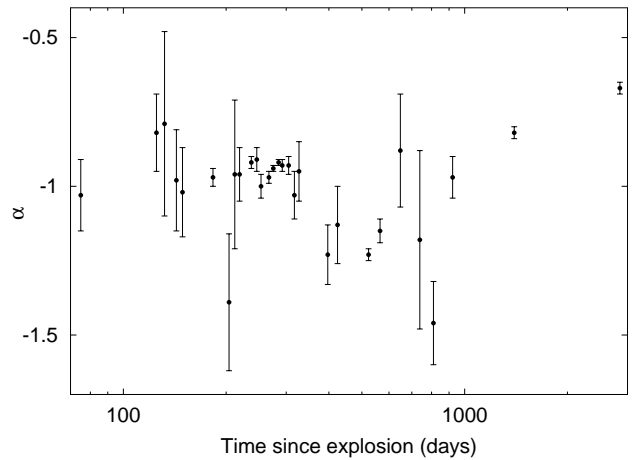}
\caption{Evolution of the observed spectral index, $\alpha$, for SN1993J, 
as obtained from fitting the radio continuum data at each epoch 
to model 2 (see text for details).}
\label{fig:alpha}
\end{figure}

Fig. \ref{fig:alpha} shows the evolution of the observed spectral
index of SN1993J from $\sim$70 up to 2820 days after the explosion.
For each epoch, we fitted the available radio continuum data 
(Van Dyk, private communication) using model 2 as described above.
While at early epochs the fits are not superb, 
they seem to be compatible with $\alpha \approx -1$. 
However, for epochs $\gtrsim$1000 days, there seems to be a clear trend: 
the observed spectral index, $\alpha$, becomes progressively less steep.
This spectral evolution could be explained assuming significant energy losses, 
mainly associated with synchrotron, as well as Coulomb and expansion losses. 
Fransson and Bj\"ornsson  (\cite{fb98}) proposed a model 
that reproduces reasonably well the observed spectral evolution 
of SN1993J. 
In their model, a constant fraction of the shocked thermal electrons 
--characterized by a constant spectral index-- are injected and accelerated. 
These electrons lose their energy mainly due to synchrotron losses, 
thus steepening the integrated electron spectrum. 
Moreover, these authors predict that, for the physical 
parameters they are considering, radiative (synchrotron) losses 
start to become less important at $t \geq$ 1000 days at frequencies below
$\sim$2.3 GHz, whereas energy losses of the electrons due to 
adiabatic and Coulomb cooling contribute significantly below 2.3 GHz, 
which seems consistent with the observational results.
We note that a multi-frequency monitoring of SN1993J at radio frequencies down to $\sim 70$ MHz, 
would likely prove to be very useful to further constrain the population of electrons 
responsible for the emission, as well as to better understand the late-time 
radio evolution of SN1993J. 

\begin{figure}
 \centering
 \includegraphics[angle=-90,width=9cm]{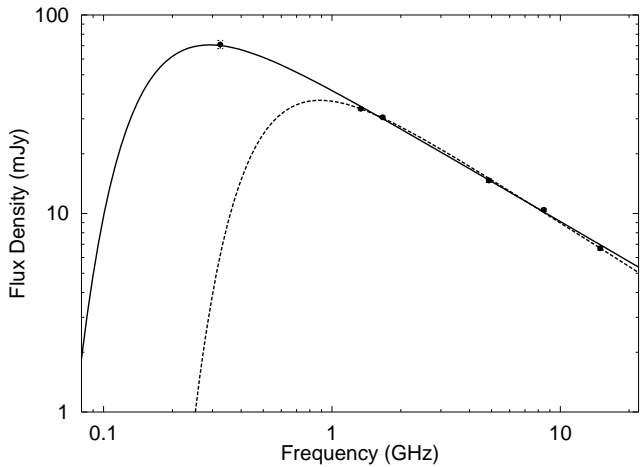}
 \caption{
Fits to the continuum radio spectrum of SN1993J on 17 and 21 December 2000. 
Although both fits correspond to model 2,  in one case (solid line) all radio measurements
are used, while in the other case (dashed line) all {\it but} the P-band data point are used. 
This results in a turnover frequency much higher than actually observed.
See text for details.
}
 \label{fig:fits_noP}
\end{figure}

Fig. \ref{fig:fits_noP} shows the fits of model 2 
to all data points (solid line), and 
to all but the P-band data point (dashed line). 
Because any absorption model will obviously fit
by turning down sharply below the lowest frequency data point, 
the availability of radio continuum measurements at the lowest 
possible frequencies is important,
in particular as a radio supernova ages. 
Had the P-band data not been available, we could 
have wrongly concluded that the peak of the SN1993J spectrum is 
at $\nu \sim$ 0.8 GHz at a level of $\sim$35 mJy. 
Having the P-data point available, we conclude
that the peak is around, or below 0.3 GHz, 
and at least at a level of 70 mJy.
(We note that a similar fit to all data, but using model 3 rather than model 2, 
would  not change this result.)
Similarly, the best fit to our SN1993J radio data using models 2 or 3 
as described above, but not including the P-band measurement,
yields $\alpha = - 0.73$, a somewhat steeper 
value than is obtained if all data points are used, $\alpha = - 0.67$, 
but within 2$\sigma$ of it.

\section{Summary}
\label{summary}

We present VLA radio continuum measurements of SN1993J in M81
taken on 17 and 21 December, 2000, about 2820 days after the supernova
explosion,  spanning the frequency range from 0.32 to 14.9 GHz.
We discuss in detail the P-band (322-327 MHz) observations, since 
the task of obtaining the flux density at these frequencies is much harder
than at higher frequencies, due to confusion from many nearby and far away, 
strong sources, e.g. M82.
We also point out the importance of having a radio spectral coverage
that extends down to P-band as the supernova ages, since otherwise the 
conclusions can be wrong.

We discuss three models for the late radio emission of SN1993J: 
(i) a simple power-law spectrum, 
(ii) a power-law spectrum 
free-free absorbed by a screen of homogeneously distributed ionized gas, 
and (iii) a power-law spectrum free-free absorbed by a ``clumpy'' medium. 
We find that the best fit to the data is yielded by a power-law spectrum
free-free absorbed by either a homogeneous, or a clumpy, distribution of ionized gas. 
However, a mixture of homogeneous and
clumpy absorbers, while not preferred, cannot be totally ruled out.
The radio spectrum between 0.32 and 14.94 GHz
is well characterized by $\alpha = -0.67 \pm 0.02 $ ($S_\nu \propto \nu^\alpha$),
typical of an optically thin radio supernova, and 
is significantly different from that obtained at epochs $\lessim$1000 days
between 1.4 and 14.9 GHz, which indicates an ongoing spectral evolution 
in the radio emission from SN1993J.
Since the spectral index $\gamma$ of the electron distribution 
($N_E=N_0\,E^{-\gamma}$),  is related to the observed spectral index 
by $\gamma = 1 - 2\,\alpha$, 
our value of $\alpha$ translates into $\gamma=2.34\pm0.04$,
which agrees well with that predicted by 
Fransson \& Bj\"ornsson (\cite{fb98}) 
for $t \gtrsim 1000$ days.
We also fit the available radio continuum data of SN1993J for
the period from $\sim$70 up to 2820 days since the explosion, 
using a power-law spectrum free-free
absorbed by a screen of homogeneously distributed ionized gas. 
The fit shows that the observed spectral index of SN1993J 
has been slowly evolving since $t \sim$1000 days on, with 
$\alpha$ increasing from a value close to -1 to -0.67. 
This spectral evolution seems to suggest that, in addition to the 
radiative (synchrotron) losses,  
adiabatic cooling and ionization (Coulomb) losses at the lowest frequencies 
might be contributing significantly to the integrated electron spectrum. 
An extension of the low frequency observations further down to $\sim$70 MHz should 
help to fine tune the physical parameters of SN1993J, and discern between 
the homogeneous or clumpy nature of its circumstellar medium. 
From a technical point of view, such observations will be challenging.

\begin{acknowledgements}
We thank Lucas Lara for his help with some aspects of the data 
reduction process, and Gianfranco Brunetti for discussions.
We thank Kurt Weiler for making the VLA data publicly available.
We are also grateful to an anonymous referee for a careful
and constructive review of our manuscript.
MAPT is grateful to the Instituto de Astrof\'{\i}sica de Andaluc\'{\i}a 
for its hospitality during a short visit. 
This research has been supported by a Marie Curie Fellowship
of the European Community (contract IHP-MCFI-99-1),
and by the Spanish DGICYT grants AYA2001-2147-C02-01
and AYA2001-2147-C02-02.
The VLA is an instrument of the National Radio Astronomy Observatory, 
which is a facility of the National Science Foundation operated under 
cooperative agreement by Associated Universities, Incorporated.

\end{acknowledgements}

\end{document}